\documentstyle[preprint,prd,floats,aps,array]{revtex}
\tighten        
\begin{document}

\input{epsf.sty}

\draft


\title{Three Dimensional Distorted Black Holes}

\author{Steven Brandt$^{(1)}$, Karen Camarda$^{(2,3)}$, Edward
  Seidel$^{(4,5,6)}$, and Ryoji Takahashi$^{(4,7)}$}

\address{
$^{(1)}$190 Drexel Ave., Lansdowne, PA 19050\\
$^{(2)}$Department of Chemical and Petroleum Engineering, University of
  Kansas,\\ Lawrence, KS 66045\\
$^{(3)}$Department of Physics and Astronomy, Washburn University,
  Topeka, KS 66621\\
$^{(4)}$Max-Planck-Institut f{\"u}r Gravitationsphysik, Am
  Muehlenberg 5, D-14476 Golm, Germany\\
$^{(5)}$ National Center for Supercomputing Applications,
Beckman Institute, 405 N. Mathews Ave., Urbana, IL 61801 \\
$^{(6)}$ Departments of Astronomy and Physics,
University of Illinois, Urbana, IL 61801 \\
$^{(7)}$ Theoretical Astrophysics Center, Juliane Maries Vej 30, 
2100 Copenhagen, Denmark \\
}

\date{\today}

\maketitle

\begin{abstract}
We present three-dimensional, {\it non-axisymmetric} distorted black
hole initial data which generalizes the axisymmetric, distorted,
non-rotating~\cite{Bernstein93a} and
rotating~\cite{Brandt94a} single black hole data
developed by Bernstein, Brandt, and Seidel.
These initial data should be useful for studying the dynamics of fully
3D, distorted black holes, such as those created by the spiraling
coalescence of two black holes.  We describe the mathematical
construction of several families of such data sets, and show how to
construct numerical solutions.  We survey quantities associated with
the numerically constructed solutions, such as ADM masses, apparent
horizons, measurements of the horizon distortion, and the maximum
possible radiation loss ($MRL$).

\end{abstract}

\draft

%
%

\section{Introduction}
\label{sec:Intro}

Black hole studies have received significant attention in
numerical relativity over the past several years, as the computers needed to
solve the
Einstein equations grow in power.  The need for computer generated
templates of gravitational waves has become more pressing since
gravitational wave detectors, such as LIGO and GEO600, are nearing completion
and should begin taking data in about a
year~\cite{Abramovici92,Hough94b}.  Binary black hole coalescence
events are considered to be prime candidates for the first detection
of gravitational waves~\cite{Flanagan97a}.  

Several theoretical approaches have been developed for treating these
systems.  So far, the post-Newtonian approximation (PN) has provided a
good understanding of the early slow adiabatic inspiral, or
``far-limit'', phase of these systems~\cite{Damour:2001bu,Damour:2000ni,Damour:2000we,Jaranowski98a,Blanchet95}.  Similarly, for the final
moments, when the black holes are close enough to each other to sit inside
a common gravitational well, one can successfully apply the ``close
limit'' approximation (CL)~\cite{Price94a}, which  effectively
describes the whole system as a perturbation of a single black hole
which rapidly ``rings-down'' to stationarity.  Before this last stage,
though, when the black holes are still close to the {\it innermost
  stable circular orbit}, the orbital dynamics are expected to
yield to a plunge and coalescence. No approximation method can be
applied in this highly nonlinear phase and it is generally expected
that one can only treat the system by a full numerical
integration of Einstein's equations.

Ideally, one would like to start a full nonlinear integration of the
Einstein equations with initial data that correspond to black holes in
the early adiabatic inspiral phase. Such data would be straight-forward
to compute~\cite{Tichy02}, and would correspond well to a
realizable astrophysical situation.  Unfortunately,
there are several technical difficulties involved in numerically
evolving such a system all the way through coalescence to ringdown.
Accordingly, in order to develop and test fully nonlinear numerical codes
while the difficulties are being resolved,
the strategy has been to evolve initial data that can be used as a model of
the merger event.  For example, initial data corresponding to holes
which are initially very close together have been
evolved~\cite{Anninos93b,Brandt00,Alcubierre00b}, as have data sets
which correspond to a single very distorted
hole~\cite{Abrahams92a,Brandt94c,Alcubierre01a}.

In this paper we describe initial data sets which correspond to
three-dimensional distorted black holes, both with rotation and
without. The formalism employed extends the axisymmetric work done
by Bernstein, Brandt, and Seidel~\cite{Bernstein93a,Brandt94a}.
Although these data sets and their evolution have already been
described briefly in several places~\cite{Alcubierre01a,Brandt97c,Allen97a,Allen98a,Baker99a,Alcubierre01b,Camarda97a}, 
in this paper, we provide a more complete description of them.

The paper is organized as follows. In
Section~\ref{sec:initdata} we describe the mathematical setup and
numerical construction of the data sets.
In Section~\ref{sec:survey} 
we give results of parameter studies
of their properties, such as ADM masses, apparent
horizon positions, distortion measures, and maximal radiation loss.
Then, we conclude our results in Section~\ref{sec:conc}.

%
%

\section{CONSTRUCTING DISTORTED 3D BLACK HOLES}
\label{sec:initdata}

The standard approach to constructing initial data for Einstein's
equations is to consider a 3+1 split (3 space and 1 time) of the full
four-dimensional theory.  In this way the equations divide naturally into
two classes:  four constraint equations, and six evolution
equations.  The procedure for performing the split is described extensively
in many places, notably in the review article by York~\cite{York79}.  The
constraints, which are elliptic in nature, relate
the 3-metric $\gamma_{ij}$ and extrinsic curvature $K_{ij}$ at any
coordinate time $t$.  In our case, we solve the
constraints to obtain data sets for some initial time labeled by
$t=0$.  The constraints can be further split into the Hamiltonian
constraint and the momentum constraints.  Here, we consider vacuum spacetimes,
for which the Hamiltonian constraint is
\begin{equation}
R+(\mbox{tr}K)^2-K^{ij}K_{ij} = 0,
\label{eqn:ham}
\end{equation}
and the three momentum constraints are
\begin{equation}
D_i (K^{ij}-\gamma^{ij}\mbox{tr}K) = 0,
\label{eqn:mom}
\end{equation}
where $R$ and $D_i$ are the scalar curvature and covariant derivative
associated with the 3-metric $\gamma_{ij}$.

To solve the constraints, it is common to use York's conformal
decomposition method~\cite{York79}.  This method starts by factoring
out a function $\psi$, known as the conformal factor, from the
3-metric and extrinsic curvature tensor components in the following way:
\begin{eqnarray}
\gamma_{ij} & = & \psi^4 \hat{\gamma}_{ij}, \\
K_{ij} & = & \psi^{-2} \hat{K}_{ij}.
\end{eqnarray}
If we use this decomposition, and restrict ourselves to initial slices
with vanishing $\mbox{tr}K = \gamma^{ij}K_{ij}$, the constraint equations
take the following form:
\begin{eqnarray}
\hat{\Delta} \psi & = & \frac{1}{8} \psi \hat{R} - \frac{1}{8}
\psi^{-7} \hat{K}^{ij}\hat{K}_{ij} \\
\hat{D}_i \hat{K}^{ij} = \partial_{i} \hat{K}^{ij}&+&\hat{\Gamma}^i_{ik} \hat{K}^{kj} +
\hat{\Gamma}^{j}_{ik}\hat{K}^{ik} = 0,
\end{eqnarray}
where $\hat{\Delta}$ is the Laplacian, $\hat{R}$ the scalar curvature, 
and $\hat{\Gamma}^{j}_{ik}$ the Christoffel symbol associated with the 
conformal 3-metric $\hat{\gamma}_{ij}$.  Note that the conformal
factor does not appear in the momentum constraint equations.  This
allows us to use the following procedure to derive initial data.
First, one specifies the conformal 3-metric $\hat{\gamma}_{ij}$
freely.  Then, the momentum constraint equations are solved for the
conformal extrinsic curvature.  Finally, one solves the Hamiltonian
constraint for the conformal factor.  This is the procedure we
use to create the 3D black hole initial data sets described below.

\subsection{Distorted non-rotating black hole}
\label{subsec:dbh}

As part of his thesis work, Bernstein studied initial data sets
corresponding to single black holes which were
non-rotating, distorted, and axisymmetric~\cite{Bernstein93a}. For these
non-rotating black holes, the extrinsic curvature was taken to vanish.
In this case,
the vacuum momentum constraint equations are satisfied identically, and the
Hamiltonian constraint reduces to
\begin{equation}
\hat \Delta \psi = \frac{1}{8}\psi \hat R.
\label{eqn:conformal_hamiltonian}
\end{equation}
The form of the conformal 3-metric was that used by Brill in his
study of pure gravitational wave spacetimes~\cite{Brill59}.
Using spherical-polar coordinates, one can write the 3-metric as
\begin{equation}
\label{eqn:sph-cood}
dl^2 = \psi^4 [e^{2q} (dr^2 + r^2 d \theta^2) + r^2 \sin^2 \theta d
\phi^2],
\end{equation}
where $q$ is the Brill ``packet'' which takes some functional form.
Using this ansatz with~(\ref{eqn:conformal_hamiltonian}) leads to
an elliptic equation for $\psi$ which must be solved
numerically. The inner boundary of our computational domain is the throat
of the black hole, and there we apply an isometry condition on $\psi$
which maps the region exterior to the throat to another asymptotically flat
region interior to it.
At the outer boundary, we apply the Robin 
condition which requires $\psi \sim O(r^{-1})$ as $r \rightarrow
\infty$ is used as the outer boundary condition.  The choice of $q=0$
produces the spherically symmetric Schwarzschild solution in isotropic
coordinates.  In Bernstein's work, a logarithmic radial coordinate
$\eta$ is used, related to the isotropic radial coordinate $r$ by
\begin{equation}
\label{eta_coord}
\eta = \mbox{ln} (\frac{2r}{M_0}),
\end{equation}
where $M_0$ is the mass of the Schwarzschild black hole that results from
setting $q=0$. In this coordinate system, the 3-metric is
\begin{equation}
\label{eqn:metric_brill_eta}
dl^2 = \tilde{\psi}^4 [e^{2q} (d\eta^2+d\theta^2)+\sin^2 \theta
d\phi^2],
\end{equation}
and the Schwarzschild solution is 
\begin{equation}
\label{eqn:psi}
\tilde{\psi} = \sqrt{2M_0} \cosh (\frac{\eta}{2}).
\end{equation}
Note that the conformal factor in the $\eta-$coordinate system,
$\tilde{\psi}$, differs from that in the isotropic radial coordinate system
by a factor of $r^{1/2}$.

The Hamiltonian constraint in this coordinate system is
\begin{equation}
\label{eqn:hamcos_eta}
\frac{\partial^2 \tilde{\psi}}{\partial \eta^2} + \frac{\partial^2
  \tilde{\psi}}{\partial \theta^2} + \cot \theta \frac{\partial
  \tilde{\psi}}{\partial \theta} = - \frac{1}{4} \tilde{\psi}
(\frac{\partial^2 q}{\partial \eta^2} + \frac{\partial^2 q}{\partial
  \theta^2} -1).
\end{equation}

The throat of the black hole is located at $r=M_0/2$, or $\eta = 0$.
The isometry condition across the throat in the $\eta$ coordinate system
takes the particular simple form $\gamma_{ij}(\eta) =
\pm\gamma_{ij}(-\eta)$, which for the conformal factor becomes
$\tilde\psi(\eta)=\tilde\psi(-\eta)$. At the outer boundary we require that
$\tilde\psi$ have the same behavior as the spherically symmetric solution.

The Brill wave function $q$ was chosen to have the following form,
which obeys the boundary conditions but is otherwise arbitrary.
Specifically Brill showed that in order for the mass of the
hypersurface to be well defined, $q$ must vanish on the axis and decrease
radially at least as fast as $e^{-2\eta}$.  In this paper, we choose
$q$ to be of the form used by Bernstein:
\begin{equation}
q(\eta, \theta) = Q_0 \sin^n \theta\left[
    e^{-\left(\frac{\eta + \eta_0}{\sigma}\right)^2} +
    e^{-\left(\frac{\eta - \eta_0}{\sigma}\right)^2} \right].
\end{equation}
Roughly speaking, $Q_0$ is the amplitude of Brill wave,
$\eta_0$ its radial location, and $\sigma$ its width.
Regularity along the axis requires that the exponent $n$ must be
even.  We characterize initial data sets by the parameters
$(Q_0,\eta_0,\sigma, n)$.  

Note that an even exponent $n$ results in data that have
equatorial plane symmetry, so that only $\theta$ in the range $0 \le
\theta \le \pi/2$ (or equivalently, $\pi/2 \le \theta \le \pi$) need
to be considered. 

We chose to extend Bernstein's work to 3D by multiplying the
Brill wave function $q$ by a factor which has azimuthal dependence.
The particular form of this factor was $1+c \cos^2 \phi$, giving the
$q$-function
\begin{equation}
q(\eta,\theta,\phi) = Q_0 \sin^n \theta \left[
   e^{-(\frac{\eta + \eta_0}{\sigma})^2} +
   e^{-(\frac{\eta - \eta_0}{\sigma})^2}\right]
   \left( 1 + c \cos^2 \phi\right).
\label{eqn:qfunc}
\end{equation}
Here, one can see that the above axisymmetric case is recovered by
setting $c=0$.  Now the Hamiltonian constraint takes on the more
complicated form
\begin{eqnarray}
\label{eqn:ham_3d}
\frac{\partial^2 \tilde{\psi}}{\partial \eta^2}+\frac{\partial^2
  \tilde{\psi}}{\partial \theta^2} + \cot \theta \frac{\partial
  \tilde{\psi}}{\partial \theta}+\csc^2 \theta e^{2q} \frac{\partial^2
  \tilde{\psi}}{\partial \phi^2} + 2 \csc^2 \theta e^{2q}
  \frac{\partial q}{\partial \phi} \frac{\partial \tilde{\psi}}{\partial \phi}
  = \nonumber \\
-\frac{1}{4} \tilde{\psi}
  \left[\frac{\partial^2 q}{\partial \eta^2} + \frac{\partial^2 q}{\partial
  \theta^2} + 2 e^{2q} \csc^2 \theta \frac{\partial^2 q}{\partial
  \phi^2}+3 e^{2q} \csc^2 \theta \left(\frac{\partial q}{\partial \phi}\right)^2
  - 1\right]. 
\end{eqnarray}
In this case, we characterize our initial data sets by the parameters
$(Q_0, \eta_0, \sigma, c, n)$.

Notice that we still have equatorial plane symmetry with this $q$.  We
also have sufficient symmetry in the azimuthal angle $\phi$ to
restrict ourselves to the range $0 \le \phi \le \pi/2$.  This allows
us to perform computations in only one octant, resulting in a great
savings in computational time.  However, for consistency
with the non-axisymmetric rotating cases in which one cannot apply any
symmetries, we solve it on a full grid domain.  Note also that
although setting $n=2$ is valid in axisymmetry, when $c\ne0$, $n$ must
be at least 4 for the right hand side of (\ref{eqn:ham_3d}) to be
regular.  Using the same boundary conditions as above, we can solve
(\ref{eqn:ham_3d}) for the conformal factor $\tilde{\psi}$ numerically
on this spherical grid.

In order to evolve these initial data, we interpolate the conformal factor
$\tilde\psi$ and its derivatives onto a Cartesian grid. As a test of our
initial data solver, we compute the residual of the Hamiltonian constraint,
$H$, on this Cartesian grid at various resolutions. We expect to see second
order convergence in the grid spacing.

\begin{figure}[t]
\epsfxsize=10cm
\epsfysize=10cm
\hspace{3cm}
\epsfbox{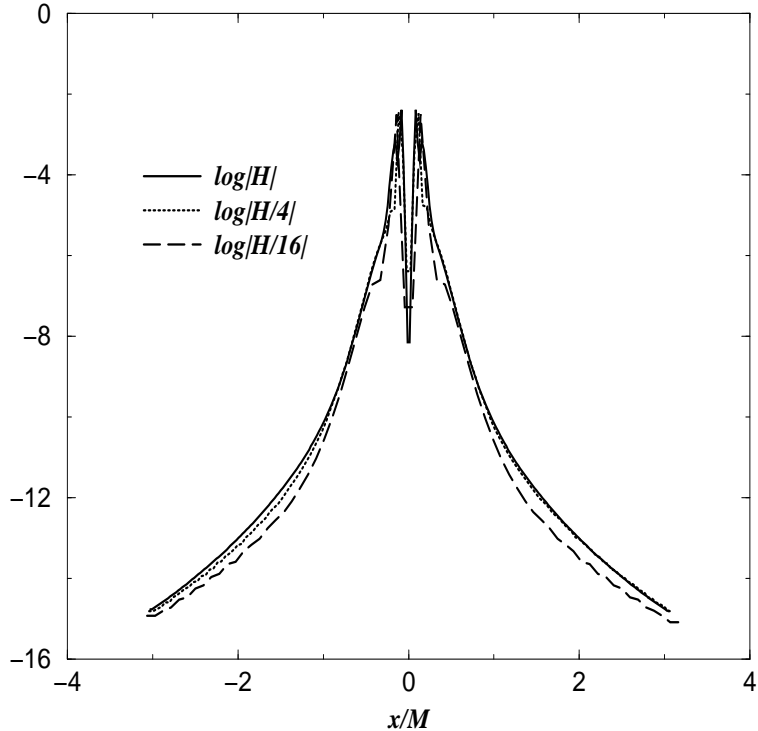} 
\caption{We show the convergence of the residual of the Hamiltonian
  constraint for distorted non-rotating black hole initial data with
  parameter set $(Q_0,\eta_0,\sigma,c,n)=(-0.5,0,1,1,4)$.  $H$ is
  plotted for Cartesian grid sizes (resolutions) $259^3 (0.024M)$,
  $131^3(0.047M)$, and $67^3(0.095M)$.  As indicated, values computed
  at lower resolutions are rescaled such that they would coincide for
  perfect second order convergence.}
 \label{fig:dbh_ham_log.conv}
\end{figure}

Fig.~\ref{fig:dbh_ham_log.conv} shows the logarithm of the residual
of the Hamiltonian constraint, $H$, on the $x$-axis at three different
resolutions for the data set
$(Q_0,\eta_0,\sigma,c,n)=(-0.5,0,1,1,4)$.  The values of $H$ are
rescaled so that they would coincide for perfect second order
convergence. We obtain near second order convergence, although some
error is seen for the coarsest resolution.

This family of initial data sets of isometric
embedding~\cite{Bondarescu02}, and their evolution as perturbations
of the Schwarzschild black hole~\cite{Allen97a,Allen98a} have been
studied.  In Section~\ref{sec:survey}, we study various physical
properties of these black holes as a function of the parameters of
this initial data.

\subsection{Distorted rotating black hole}
\label{subsec:rdbh}

Brandt and Seidel extended the above non-rotating axisymmetric initial
data sets of Bernstein to include
rotation~\cite{Brandt94a,Brandt94c,Brandt94b}.  In the same way that
Bernstein's data sets correspond to a Schwarzschild black hole
surrounded by a gravitational wave, their data sets correspond to
either a Kerr or Bowen and York black hole surrounded by a
gravitational wave.  For the rotating case, the 3-metric takes the form
\begin{equation}
dl^2 = \tilde\psi^4 \left[e^{2(q-q_0)}(d\eta^2+d\theta^2)+\sin \theta^2d\phi^2
       \right].
\label{eq:2d_rot_confmet}
\end{equation}
Note that a function $q_0$ has been subtracted from the Brill wave function
$q$. The form of $q_0$ is found by setting the 3-metric equal to the spatial
part of the Kerr metric and setting $q=0$.  Apart from
the function $q_0$ this is the same three-metric used for the non-rotating,
distorted black hole spacetime.  Here the
logarithmic coordinate $\eta$ is related to the Boyer-Lindquist
coordinate $r$ by
\begin{equation}
r = r_{+} \cosh^2 (\eta/2)-r_{-}\sinh^2(\eta/2),
\end{equation}
where $r_{\pm} = M\pm\sqrt{M^2-a^2}$.  When considered together with
the extrinsic curvature, this 3-metric allows for a distinct class of
initial data: distorted Kerr or Bowen and York black holes.  To
obtain initial data for a distorted Kerr black hole, one can set the
conformal extrinsic curvature tensor $\hat{K}_{ij}$ to that which
would be obtained for the standard Kerr metric, as shown
in~\cite{Brandt94a}.

In order to obtain a distorted Bowen and York black hole, one can use
the Bowen and York expression for the conformal extrinsic
curvature~\cite{Bowen80}:
\begin{equation}
  \hat{K}_{ij} = \left(\begin{array}{ccc}
      0 & 0 & 3J \sin^2 \theta \\
      0 & 0 & 0 \\
      3J \sin^2 \theta & 0 & 0 
    \end{array}
  \right), \\
\end{equation}
where $J$ is the total angular momentum of
spacetime\footnote{$J$ is both a free parameter and the total angular momentum
  (ADM angular momentum) of the initial data, and hence of the space
  time. This is always true. It does not matter whether the data
  is axisymmetric or not, because the total angular momentum is
  defined at infinity (i.e. it is not local).  Notice that when
  there are symmetries, it is possible to define not only the angular
  momentum at infinity, but also a quasi-local angular momentum by
  integrating the Killing vector (and both coincide).  However, even
  this quasi-local definition is possible for our case:
  for the non-axisymmetric extrinsic curvature described in this section,
  since our data is conformally flat and a maximal slice, one can use
  the conformal Killing vector to define this quasi-local angular
  momentum as described in Ref.~\cite{Dain02a}}. The conformal metric used
in this case is that resulting from setting $q_0=0$ in
(\ref{eq:2d_rot_confmet}) which is related to the conformally flat
metric.

It is not possible to extend the axisymmetric distorted Bowen and York
black holes to 3D in the same way as in the non-rotating case. This is
because if one allows the Brill wave function $q$ to depend on the
azimuthal angle $\phi$, it can be shown that the $\eta$--component of
the momentum constraint equation reduces to $J \partial_\phi q=0$; i.e.
for the Bowen and York form of the conformal extrinsic curvature, one
cannot have
both angular momentum and a Brill wave function with azimuthal dependence.
Instead, to extend the rotating case to 3D, we use the conformally flat metric
that results from setting $q=q_0$ in equation (\ref{eq:2d_rot_confmet}), and
place the $\phi$--dependence in the conformal extrinsic curvature.
The general form of
$\hat{K}_{ij}$~\footnote{There is a way to construct   ``all'' the
  solutions in an explicit form: namely construct all the   solutions
  in terms of derivatives of two free functions as in Theorem
  14~\cite{Dain01a} } used is
\begin{eqnarray}
\label{eqn:ex_mom1}
\hat{K}_{ij} & = & m \sin(m \phi) \left(
                                \begin{array}{ccc}
                                h_{0} & h_{1} &0 \\
                                h_{1} &h_{x}-h_{0}&0 \\
                                0 & 0 &-h_{x} \sin^{2}\theta 
                                \end{array}
                                \right)
               + \left(
                        \begin{array}{ccc}
                        0 & 0 & \hat{K}_{\eta \phi} \\
                        0 & 0 & \hat{K}_{\theta \phi} \\
                        \hat{K}_{\eta \phi}&\hat{K}_{\theta \phi}&0
                        \end{array}
                        \right),
\end{eqnarray}
where
\begin{eqnarray}
\label{eqn:ex_mom2}
\hat{K}_{\eta \phi} &  = & \sin^{2} \theta [3J + \cos(m\phi)
\frac{1}{\sin^{3}\theta}\partial_{\theta}(\sin^{4}\theta g)], \\
\label{eqn:ex_mom3}
\hat{K}_{\theta \phi} & = & \cos(m \phi)\sin\theta(m^{2}v - \sin^2 \theta
\partial_{\eta} g).
\end{eqnarray}
Note that we have made all $\phi$-dependence in the problem
explicit.  The functions $v, g, h_0, h_x,$ and
$h_1$ that appear above depend only $\eta$ and $\theta$.  Below we outline
a procedure for finding forms for these functions that result in a
conformal extrinsic curvature that satisfies the momentum constraint.
The quantity $J$ is a constant.

We choose the functions $h_0$ and $h_1$ to have the form
\begin{eqnarray}
h_0 & = & \partial_{\eta} \Omega - \frac{1}{\sin \theta}
\partial_{\theta} (\sin \theta \Lambda), \\
h_1 & = & \partial_{\eta} \Lambda + \frac{1}{\sin \theta}
\partial_{\theta} (\sin \theta \Omega).
\end{eqnarray}
If $m=0$, the momentum constraint equations are satisfied trivially. If
$m \ne 0$, the $\eta$, $\theta$, and $\phi$ component of momentum
constraints become, respectively,
\begin{eqnarray}
\label{eqn:moment_eta}
(\partial^2_{\eta} -1 + \partial^2_{\theta} + 2 \cot \theta
\partial_{\theta}) \Omega & = & (4 \cos \theta + \sin \theta
\partial_{\theta}) g, \\
\label{eqn:moment_theta}
(\partial^2_\eta -1 + \partial^2_{\theta} + 2 \cot \theta
\partial_{\theta}) \Lambda & = & - \csc \theta m^2 v -
\partial_{\theta} h_x - 2 \cot \theta h_x + \sin \theta
\partial_{\eta} g, \\
\label{eqn:moment_phi}
h_x + 2 \cos \theta v + \sin \theta \partial_{\theta} v & = & 0.
\end{eqnarray}

Now we expand all functions in trigonometric functions of $\theta$ to the
lowest order at
which everything can be solved.  After some calculations, we found the
minimum collection of terms for $g$ and $\Omega$ to be
\begin{eqnarray}
g & = & (\tilde{g}_0 + \tilde{g}_2 \sin^2 \theta ) \cos \theta \\
\Omega & = & \sin^2\theta \,\tilde{\Omega}_2 + \sin^4\theta
\,\tilde{\Omega}_4
\end{eqnarray}
Note that for this discussion, tildes are used to indicate functions which
depend on $\eta$ alone.

One substitutes these equations into the $\eta-$component of the momentum
constraint~(\ref{eqn:moment_eta}).  By doing this, one obtains an
equation which can be written
\begin{equation}
0 = \tilde{E}_0 + \tilde{E}_2 \cos \left(2 \theta\right) +
    \tilde{E}_4 \cos \left(4 \theta\right).
\end{equation}
Obviously, each coefficient function $\tilde{E}_i$ must vanish. Setting
$\tilde{E}_4$ to zero gives
\begin{equation}
\tilde{g}_2 = \frac{1}{7}(25\tilde{\Omega}_4 - \partial^2_{\eta}
\tilde{\Omega}_4).
\end{equation}
Setting $\tilde{E}_0 + \tilde{E}_2 = 0$ gives
\begin{equation}
\tilde{g}_0 = \frac{3}{2} \tilde{\Omega}_2.
\end{equation}
Next, we introduce an arbitrary function $\tilde y$, and redefine
$\tilde\Omega_2$ to be
\begin{equation}
\tilde{\Omega}_2 = \tilde{y} - \frac{6}{7} \tilde{\Omega}_4.
\end{equation}
Using the relation $\tilde{E}_2 - \tilde{E}_0 = 0$ we obtain
\begin{equation}
\tilde{\Omega}_4 = \frac{7}{2} (2 \partial^2_{\eta} \tilde{y} - 3
\tilde{y}).
\end{equation}
At this point, we have used the $\eta-$component of the momentum constraint
(\ref{eqn:moment_eta}) to define the functions $\Omega$ and $g$ in terms of
the arbitrary function $\tilde y(\eta)$.

In order to solve the $\theta-$component of the momentum constraint
(\ref{eqn:moment_theta}), we expand the functions $\Lambda$ and $v$ as follows:
\begin{eqnarray}
\Lambda & = & \sin \left(2 \theta\right) \tilde{\Lambda}_2 +
              \sin \left(4 \theta\right) \tilde{\Lambda}_4, \\
v & = & \tilde{v} \left[\cos \theta - \cos \left(3 \theta\right)\right].
\end{eqnarray}
Plugging these expansions into the $\theta-$component of the momentum
constraint, and using the $\phi-$component of the momentum constraint to
write $h_x$ in terms of $v$, results in an equation which can be written
\begin{equation}
0 = \tilde{L}_1 \cos \theta + \tilde{L}_3 \cos \left(3 \theta\right) +
    \tilde{L}_5 \cos \left(5 \theta\right).
\end{equation}
Clearly, it must be the case that $\tilde L_1 = \tilde L_3 = \tilde L_5 = 0$.
Setting $\tilde L_1 + \tilde L_3 + \tilde L_5 = 0$ gives
\begin{equation}
\tilde{\Lambda}_2 = -2 \tilde{\Lambda}_4.
\end{equation}
In order to solve for $\tilde{\Lambda}_4$, we set
$\tilde{L}_1/32 + \tilde{L}_3/48 = 0$,  which eliminates the
derivatives of $\tilde{\Lambda}_4$. This gives
\begin{equation}
\tilde{\Lambda}_4 = \frac{4(m^4 -16) \tilde{v} + 15 \partial_{\eta}
  \tilde{y} - 9 \partial^3_{\eta} \tilde{y}}{96}.
\end{equation}
The next step is to define $\tilde{v}$ and $\tilde{y}$ in terms of an
arbitrary function $\tilde{x}$:
\begin{eqnarray}
\tilde{v} & = & \sum_{n = 1,3,5} a_n \partial^{n}_{\eta} \tilde{x}, \\
\tilde{y} & = & \tilde{x} + a_7 \partial^2_{\eta} \tilde{x}.
\end{eqnarray}
Finally, setting $\tilde{L}_{5}=0$ and using the above substitutions produces an
equation which can be written
\begin{equation}
A_1 \, \partial_\eta \tilde x + A_3 \, \partial^3_\eta \tilde x +
A_5 \, \partial^5_\eta \tilde x + A_7 \, \partial^7_\eta \tilde x = 0
\end{equation}
Solving the linear system that results from setting each $A_i$ to zero gives
\begin{eqnarray}
a_1 & = & \frac{-15}{32 - 20 m^2}, \\
a_3 & = & \frac{39}{80 - 50m^2}, \\
a_5 & = & \frac{3}{20 (-8+5m^2)}, \\
a_7 & = & \frac{16 - m^2}{-40 + 25m^2}.
\end{eqnarray}
We now have analytic forms for the equations $v$, $g$, $h_0$, $h_1$, and $h_x$
which are functions of an arbitrary function $\tilde x(\eta)$ and which
satisfy the momentum constraint equations. In this work, we choose the
function $\tilde x$ to be the Brill wave function without the $\sin^n\theta$
factor:
\begin{equation} 
\tilde{x} = Q_0[e^{-(\frac{\eta -\eta_0}{\sigma})^2} +
e^{-(\frac{\eta+\eta_0}{\sigma})^2}].
\label{eqn:xtil}
\end{equation}
Note that if we set the
parameter $m$ to zero, the problem reduces to an axisymmetric problem,
although not to the one studied by Brandt and Seidel.  If we
further set $Q_0=0$, we recover the Bowen and York solution.

Using the conformal extrinsic curvature derived above, we can solve the
Hamiltonian constraint numerically for the conformal factor $\tilde\psi$:
\begin{equation}
\frac{\partial^2 \tilde{\psi}}{\partial \eta^2} + \frac{\partial^2
  \tilde{\psi}}{\partial \theta^2} + \cot \theta \frac{\partial
  \tilde{\psi}}{\partial \theta} + \csc^2\theta \frac{\partial^2 \tilde\psi}
  {\partial\phi^2}
  =  \frac{1}{4} \tilde{\psi} -
  \frac{1}{8} \hat{K}_{ij} \hat{K}^{ij} \tilde{\psi}^{-7}
\end{equation}
We then interpolate these data onto a 3D Cartesian grid.

\begin{figure}
\epsfxsize=10cm
\epsfysize=10cm
\hspace{3cm}
\epsfbox{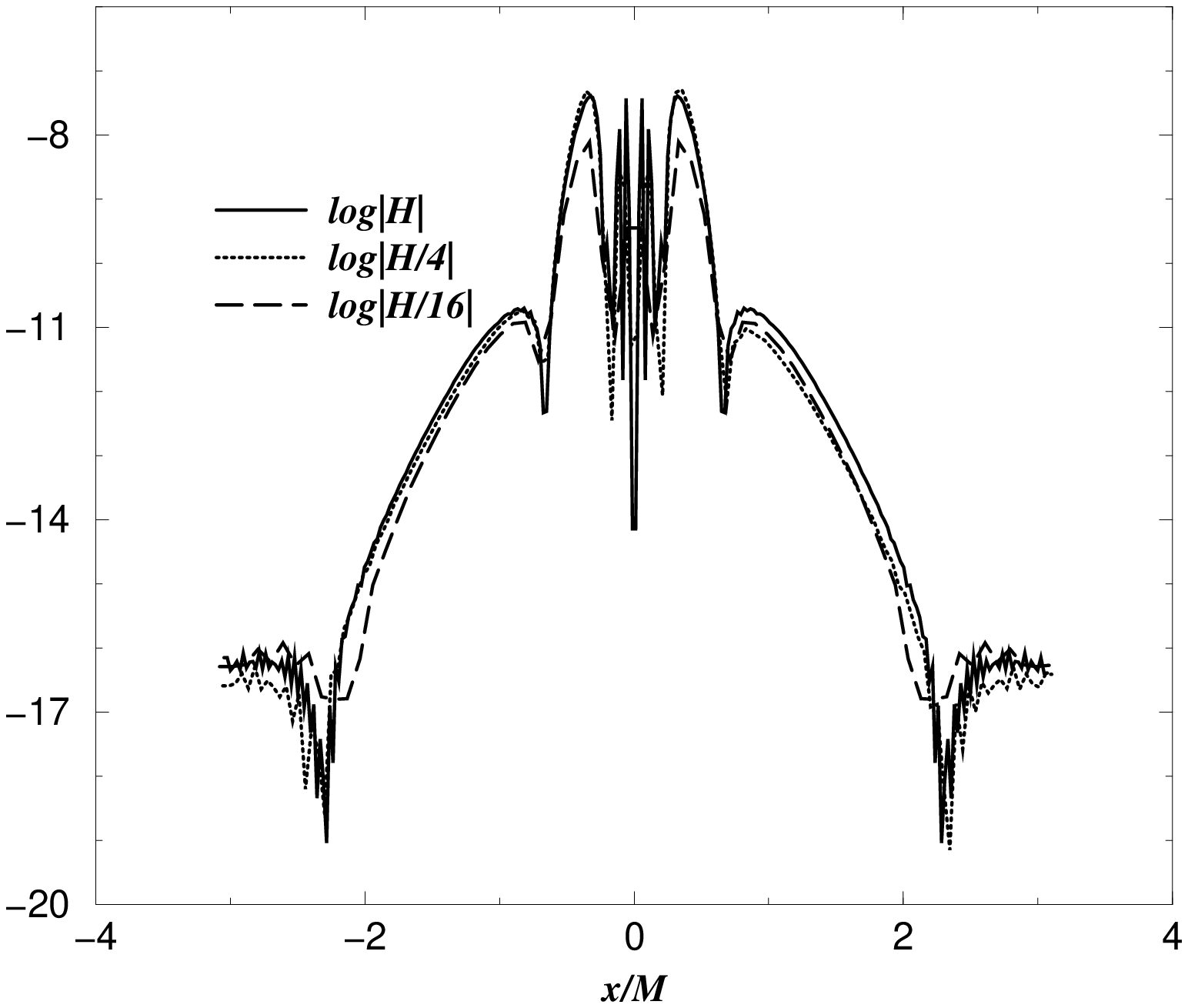} 
\caption{We show the convergence of the residual of the Hamiltonian
  constraint for distorted, rotating black hole initial data with the
  parameter set $Q_0=1.0$,$\eta_0=0$,$\sigma=1$,$J=35.0$, 
  and $m=2$.  $H$ is plotted for Cartesian grid sizes (resolutions)
  $259^3 (0.024 M)$, $131^3(0.047 M)$, and
  $67^3(0.095M)$. As indicated, values computed at lower resolutions
  are rescaled so that they would coincide for perfect second order
  convergence.}
 \label{fig:rdbh_ham_log.conv}
\end{figure}
Fig.~\ref{fig:rdbh_ham_log.conv} shows the logarithm of the residual of the
Hamiltonian constraint, $H$, on the $x-$axis at three different resolutions for
the data set $(Q_0, \eta_0, \sigma, J, m)=(1,0,1,35,2)$. The values of
$H$ are rescaled so that they would coincide for perfect second order
convergence. As in the non-rotating case, we see near second order
convergence, especially for the higher two resolutions.

%
%

\section{A SURVEY OF DISTORTED BLACK HOLE INITIAL DATA SETS}
\label{sec:survey}

In this Section we do a parameter study of the above initial data.
For all cases considered, we choose the wave location and width
parameters $\eta_0$ and $\sigma$ to be 0 and 1, respectively.  This
corresponds to a Brill wave located on the throat with unit
width.  This choice is made because we are concerned about the necessary
closeness of the outer computational boundary for the future nonlinear
evolutions of these data sets.

\subsection{Analysis of the ADM mass}
\label{subsec:adm}

Although one cannot define a local energy density of the gravitational 
field in general relativity, if the spacetime is asymptotically flat,
one can define the energy of an isolated source as measured by a
distant observer.  Arnowitt, Deser, and Misner defined an energy in
natural way for their $3+1$ formalism~\cite{Arnowitt62}.  \'O
Murchandha and York modified their expression to use the variables in
York's conformal decomposition method~\cite{Omurchadha74}.  For
conformal metrics which fall off fast enough with radius, their
expression for the ADM mass is 
\begin{equation}
M = -\frac{1}{2\pi} \oint_{\infty} D_i \psi \, d S^i.
\end{equation}
Since our numerical domains have finite extent, we
perform this integral over a sphere of large constant radius $R$,
resulting in the following expression:
\begin{equation}
M = -\frac{R^2}{2\pi} \int^{2\pi}_{0}d\phi \int^{\pi}_{0}
\frac{\partial \psi}{\partial r} \sin\theta d\theta.
\end{equation}
In the $\eta$-coordinate system, this equation is
\begin{equation}
M = -\frac{1}{2\pi} \sqrt{\frac{M_0}{2}\eta_{max}} \int^{\pi}_0 d
\phi \int^{\pi}_{0} \left.\left(\frac{\partial\tilde{\psi}}{\partial\eta} -
  \frac{\tilde\psi}{2}\right)\right|_{\eta_{max}} \sin \theta d\theta.
\label{eqn:adm_mass_eta}
\end{equation}
When we state the ADM mass for a spacetime, we have confirmed that the 
error involved in computing the integral at a finite radius is not
significant by making sure that the integral as a function of $R$ is
approaching a constant value where we are performing the
integration.

For distorted non-rotating cases, we studied a variety of
Brill wave amplitudes ($Q_0$) and azimuthal factors ($c$).
To better understand the results, we first note that the azimuthal factor
in the Brill wave function can be written
\begin{equation}
1+c \cos^2 \phi = 1+\frac{c}{2}+\frac{c}{2}\cos(2\phi).
\end{equation}
That is, it has a constant part and an oscillatory part. The constant part
by itself in effect increases the Brill wave amplitude.

\begin{figure}[t]
\epsfxsize=10cm
\epsfysize=10cm
\hspace{3cm}
\epsfbox{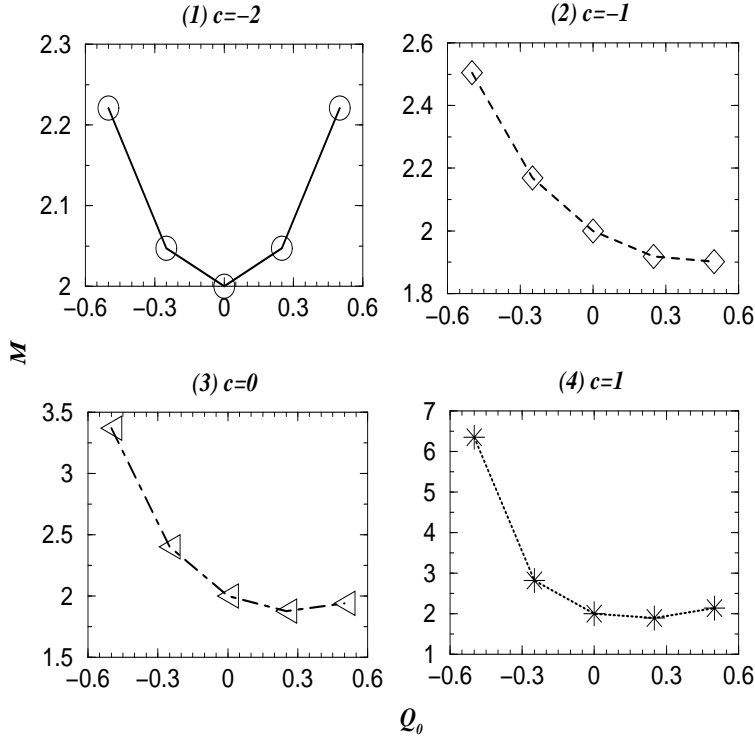} 
\caption{We show the ADM mass as a function of the Brill wave
  parameters $Q_0$ for four different values of $c$.  The wave
  location and width parameters $\eta_0$ and $\sigma$ are 0 and 1,
  respectively.  Grid size (resolution) is $131^3$ ($\Delta x=0.2$).}
 \label{fig:dbh_adm4}
\end{figure}
Fig.~\ref{fig:dbh_adm4} shows plots of ADM masses as functions of
$Q_0$ and $c$.  In panel 1 (for $c=-2$), the constant term in the
azimuthal factor vanishes.  One expects that for this case, one will
obtain the same spacetime if one changes the sign of the Brill wave
amplitude, only rotated.  Thus, the ADM mass should be symmetric
about $Q_0=0$, which is seen.  In panel 2 (for $c=-1$), we see the results
consistent
with an effective amplitude decrease.  For $Q_0<0$, the mass increases
with decreasing $Q_0$, and has lower values than the corresponding
$Q_0$ in axisymmetry.  For $Q_0>0$, the minimum in the mass occurs at
a higher value of $Q_0$ than the axisymmetric case, and it has a lower
value at any given $c$.  In panel 3 ($c=0$), the results are what one
might expect given Bernstein's results in axisymmetry, and the form of
the  azimuthal factor in the $q$-function.  Bernstein found that in
axisymmetry, when one increases $Q_0$, the ADM mass initially
decreases, then increases with $Q_0$.  Although the initial increase
seems counter-intuitive, he points out that when one increases the
amplitude of the Brill wave, it is possible to simultaneously decrease
the mass of the underlying black hole, which can give a net decrease
in the mass.  Bernstein also found that for the negative Brill wave
amplitudes, the ADM mass increases monotonically with decreasing
$Q_0$.  Our results for $c=0$ agree with the above Bernstein's
results.  Panel 4 ($c=1$) shows the same features as the panel 3.
However, the minimum in the ADM occurs at a lower value of $Q_0$ and is
larger at $Q_0=0.6$.  From Eq.~(\ref{eqn:qfunc}), the case of $Q_0=0$
corresponds to
a Schwarschild black hole. For this case, the ADM mass should be the
Schwarzschild mass $M_0$, which in this work is chosen to be 2. This is the
value seen.

The effect of the oscillatory part of the azimuthal factor by itself
can be seen in panel 1 of Fig.~\ref{fig:dbh_adm4}.  Increasing the
absolute value of $c$ increases the ADM mass.  For positive values
$c$, both the constant term and the oscillatory term are increased,
and the effect of the ADM mass has the same character as increasing
the Brill wave amplitude.  For negative values of $c$,  the situation
is more complicated.  For the range $-2 \le c \le 0$, the magnitude of 
the constant term decreases with decreasing $c$ while the magnitude
of the oscillatory term increase.  The relative effect on the ADM mass
within this range depends on the sign of the Brill wave amplitude.

For distorted rotating cases, we investigate how the ADM mass
behaves as a function of the Brill wave amplitude ($Q_0$), the
total angular momentum ($J$), and the azimuthal dependence
variable ($m$).

\begin{figure}[t]
\epsfxsize=10cm
\epsfysize=10cm
\hspace{3cm}
    \epsfbox{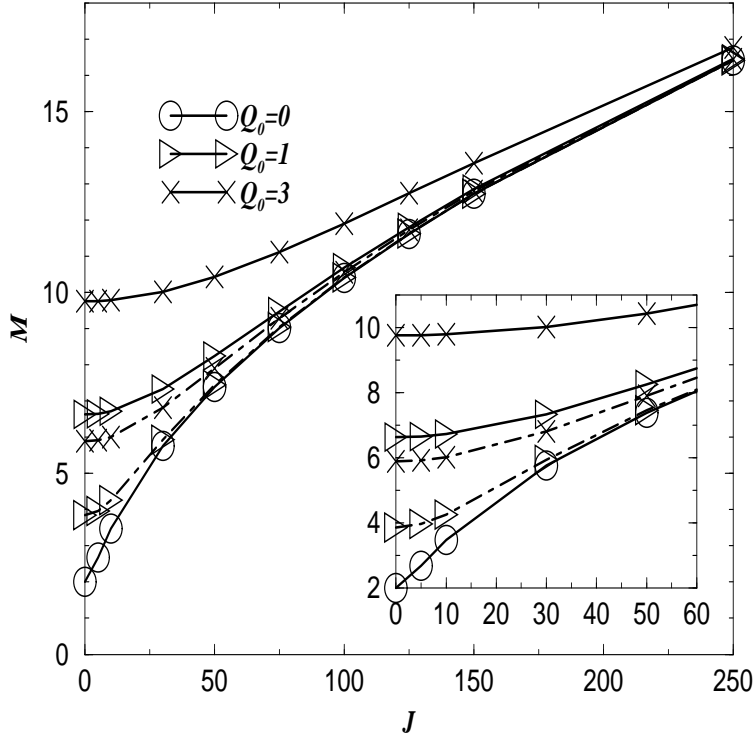} 
    \caption{We show the ADM mass $M$ as a function of the total angular
      momentum $J$ for various various values of the Brill wave amplitude
      $Q_0$. The solid
      lines show the axisymmetric cases ($m=0$), and the dashed line shows the
      non-axisymmetric cases ($m=2$).}
    \label{fig:rdbh_adm}
\end{figure}
Fig.~\ref{fig:rdbh_adm} shows the ADM mass $M$ as a function of the
total angular momentum $J$ for various values of the Brill wave amplitude
$Q_0$.  The ADM masses increase monotonically with increasing
amplitude for both axisymmetric and non-axisymmetric cases.  When the
total angular momentum $J$ is small, it is not effective for
increasing the ADM mass.

We also observe the different results when comparing axisymmetric and
non-axisymmetric cases.  In the axisymmetric cases ($m=0$), the
non-vanishing components of the conformal extrinsic curvature are
$\hat{K}_{\eta\phi}$~(\ref{eqn:ex_mom2}) and
$\hat{K}_{\theta\phi}$~(\ref{eqn:ex_mom3}).
By substituting Eq.~(\ref{eqn:xtil}) into Eqs.~(\ref{eqn:ex_mom2}), we
see that the component $\hat{K}_{\eta\phi}$ can be written
\begin{equation}
\hat{K}_{\eta\phi}(m=0) = f_0 + f_2 \cos\left(2\theta\right) + f_4
                          \cos\left(4\theta\right) + f_6 \cos\left(6\theta
                          \right)
\label{eqn:axi_ex}
\end{equation}
where
\begin{eqnarray}
f_0 &=& \frac{3}{2} J + \frac{1}{8} \tilde{g}_0 + \frac{1}{16} \tilde{g}_2 \\
f_2 &=& -\frac{3}{2} J + \frac{1}{2} \tilde{g}_0 + \frac{5}{16} \tilde{g}_2 \\
f_4 &=& -\frac{5}{8} \tilde{g}_0 - \frac{9}{16} \tilde{g}_2 \\
f_6 &=& \frac{1}{32} \tilde{g}_2
\end{eqnarray}

On the other hand, for $m=2$, this component takes the form
\begin{eqnarray}
  \hat{K}_{\eta\phi}(m=2)&=&f'_0 + f'_1 \cos\left(2\phi\right) + f'_2
  \cos(2\phi-6\theta)+f'_3
  \cos(2\phi-4\theta)+f'_4 \cos(2\phi-2\theta) + \nonumber\\
  & &f'_4 \cos(2\phi+2\theta) -f'_3 \cos(2\phi+4\theta) + f'_2
  \cos(2\phi+6 \theta)
\label{eqn:nonaxi_ex}
\end{eqnarray}
where
\begin{eqnarray}
  f'_0 &=& \frac{3}{2} J \\
  f'_1 &=& \frac{1}{8}\tilde{g}_0 \frac{1}{16} \tilde{g}_2 - \frac{3}{2} 
  J \\
  f'_2 &=& \frac{7}{64} \tilde{g}_2 \\
  f'_3 &=& \frac{5}{16} \tilde{g}_0 - \frac{9}{32} \tilde{g}_2 \\
  f'_4 &=& \frac{1}{4} \tilde{g}_0 - \frac{9}{64} \tilde{g}_2
\end{eqnarray}
If we compare the constant components of Eq.~(\ref{eqn:axi_ex}) and
Eq.~(\ref{eqn:nonaxi_ex}), we see that $f_0$ incorporates both the
total angular momentum $J$ and the amplitude $Q_0$, whereas $f'_0$
incorporates only $J$.  Because of this, for low spin, the ADM mass
increases more rapidly with amplitude for the axisymmetric cases, as
can be seen in Fig.~\ref{fig:rdbh_adm}.  As $J$ increases, we see the
effect of the Brill wave amplitude decreasing and approaching the pure
Bowen-York values.

\begin{figure}[t]
\epsfxsize=10cm
\epsfysize=10cm
\hspace{3cm}
    \epsfbox{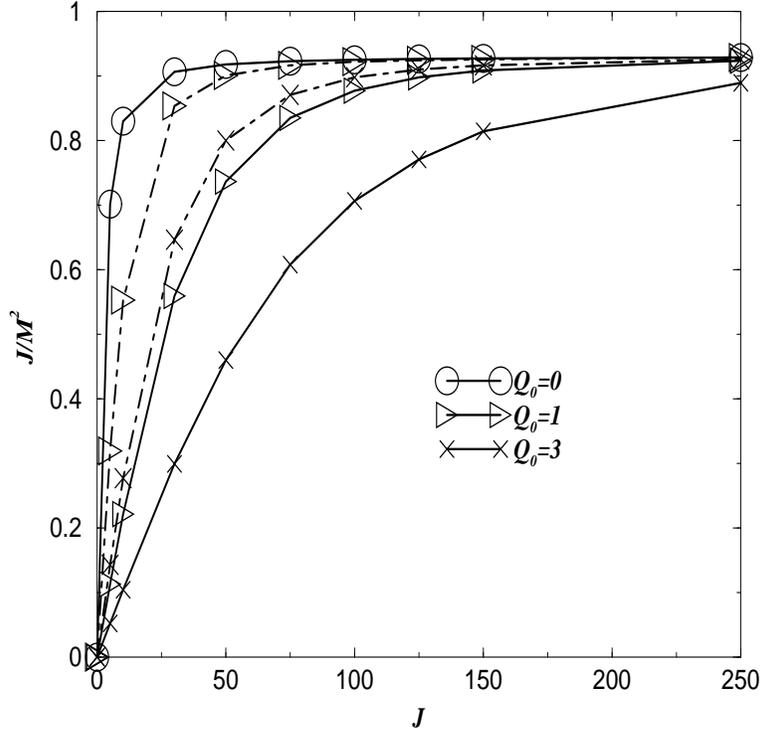} 
    \caption{We show $J/M^2$ as a function of the total angular
      momentum $J$ for various various values of the Brill wave amplitude
      $Q_0$. The solid lines show the axisymmetric cases ($m=0$), and
      the dashed line shows the non-axisymmetric cases ($m=2$).}
    \label{fig:rdbh_am}
\end{figure}
Fig.~\ref{fig:rdbh_am} shows $J/M^2$ as a function of the total
angular momentum $J$ for various values of the Brill wave amplitude
$Q_0$.  Whereas the ADM mass increases more rapidly with
amplitude for the axisymmetric cases, $J/M^2$ increases more rapidly
for non-axisymmetric cases.  Also, as $J$ increases, we
see the effect of the Brill wave amplitude decreasing and approaching
the pure Bowen-York values.  This is consistent with the
above ADM mass results.  Notice that for the Bowen-York data, $J/M^2$ does not
converge to unity as discussed in~\cite{Dain:2002ee}.  This indicates
that even for non-axisymetric cases, $J/M^2$ for our initial data will
never reach unity.  However, it is interesting to point out that if we
add a non-axisymmetric perturbation to the second fundamental form
in~\cite{Dain:2002ee}, it would cause $J/M^2$ to increase even more.

\subsection{Analysis of apparent horizons}
\label{subsec:ah}

In this Section we discuss the location and properties of the
apparent horizon (AH) in these initial data sets.  Studying the apparent
horizon gives us a detailed understanding of the shape and mass of
the black holes.

Defining $s^{\mu}$ to be the outward-pointing space-like unit normal
of a two-sphere ${\cal{S}}$ embedded in a constant time slice
$\Sigma$ with timelike unit normal $n^{\mu}$, we can construct the
outgoing null normal to any point on ${\cal{S}}$ as $k^{\mu} =
n^{\mu} + s^{\mu}$.  The surface ${\cal{S}}$ is called a marginally
trapped surface if the divergence of the outgoing null vectors
$\nabla_{\mu} k^{\mu}$ vanishes, or equivalently, if~\cite{York89}
\begin{equation}
\Theta = D_i s^i + K_{ij} s^i s^j - K = 0,
\end{equation}
where $\Theta$ is the expansion of outgoing rays.  The AH is defined
as the outer-most marginally trapped surface.

By construction, all of our data sets contain an apparent horizon.
Gibbons showed that an isometry surface, such as the throat, is an
extremal area surface~\cite{Gibbons72}. This means that in the some
cases (for example, the time symmetric cases) the throat will be an
apparent horizon. But, in general, the apparent horizon
will not be on the throat.

We have computed the coordinate location of the apparent horizon in
our initial data sets using the flow algorithm described
in~\cite{Alcubierre98b}.

\begin{figure}[t]
    \epsfxsize=10cm
    \epsfysize=10cm
    \hspace{3cm}
    \epsfbox{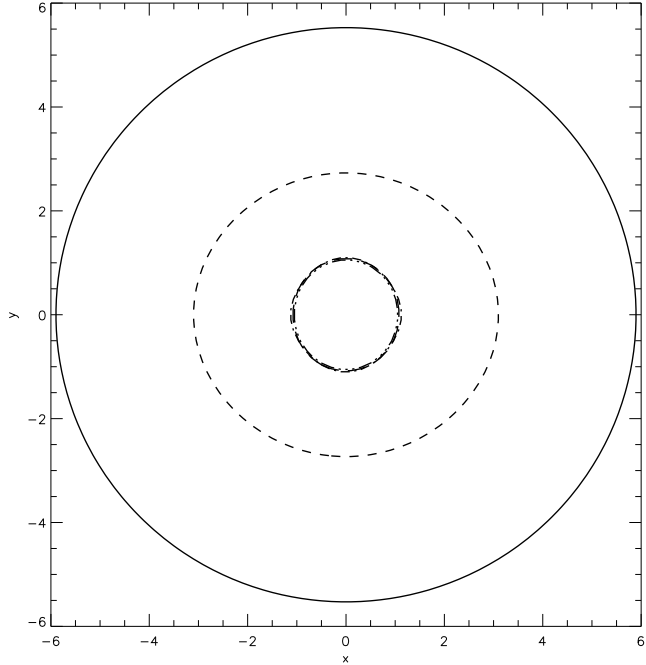} 
    \caption{The coordinate location of the apparent horizon is plotted for
      amplitudes $Q_0=(-0.6,-0.5,-0.25,0.25,0.5)$.
      The other parameters are $(\eta_0,\sigma,c)=(0,1,1)$.
      The outermost AH corresponds to $Q_0=-0.6$, and the dashed line
      corresponds to $Q_0=-0.5$. The remaining data sets have an AH at the
      inner circle, which is the throat. Grid size (resolution) is
      $131^3 (0.2)$.}
 \label{fig:dbh_ah}
\end{figure}
Fig.~\ref{fig:dbh_ah} shows the coordinate location of the apparent
horizon for several non-rotating initial data sets. The parameters are
$(Q_0,c)=(Q_0,1)$
with the following values of $Q_0$ from the outer to the inner:
$Q_0=(-0.6,-0.5,-0.25,0.25,0.5)$.  We observe that data sets with
$Q_0=(-0.25,0.25,0.5)$ have AH's which
are on the throat, while data sets with $Q_0=(-0.6,-0.5)$ form new horizons.
Although the above results are only for a few examples, we observe that
the azimuthal
factor is large enough that the effective amplitude increase results
in another minimal surface forming.

\begin{figure}[t]
\epsfxsize=10cm
\epsfysize=10cm
\hspace{3cm}
 \epsfbox{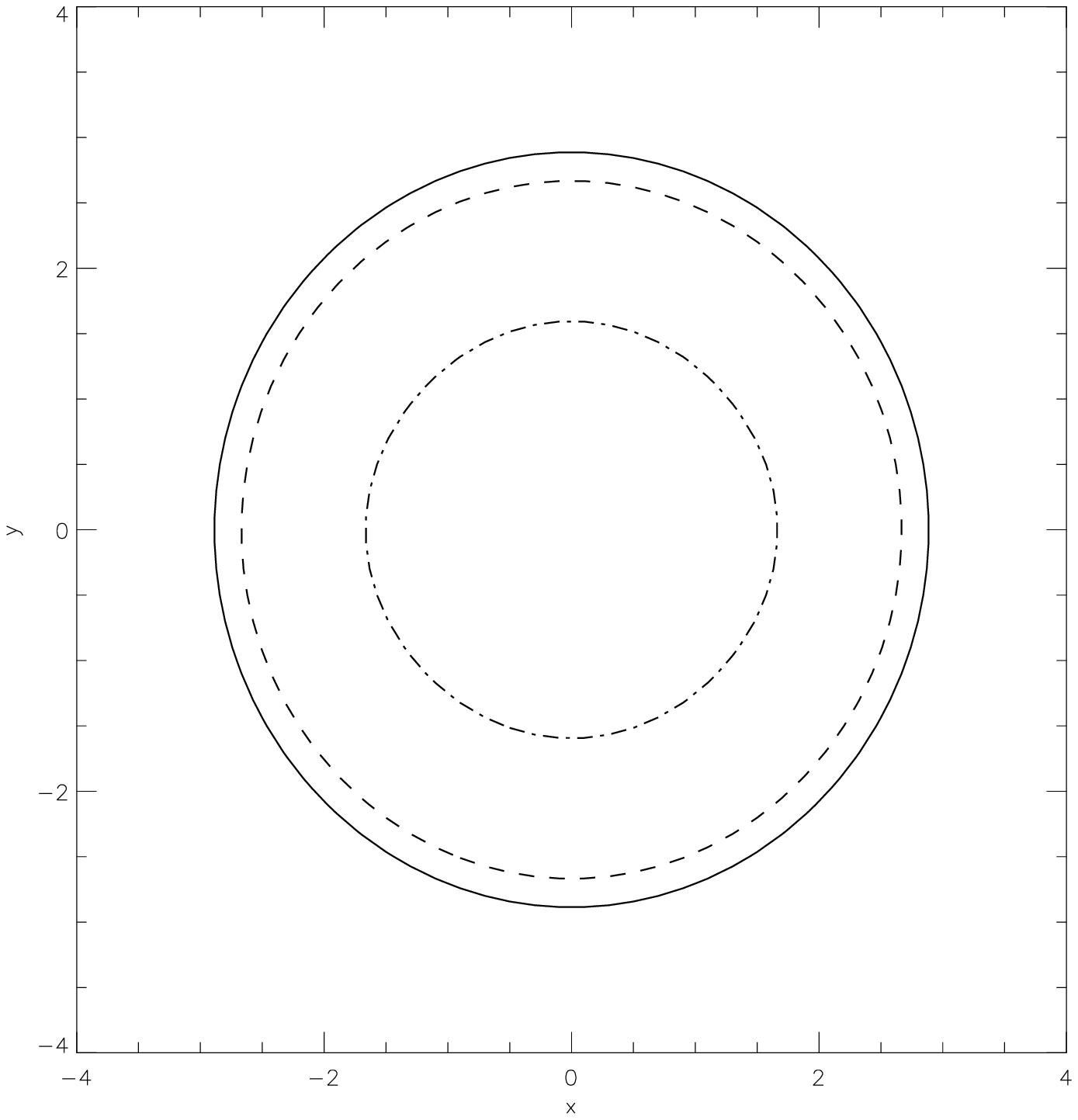}
\caption{The coordinate location of the apparent horizon of
  axisymmetric cases for various total angular momentum parameter $J$
  :$150, 30$,and $0$ which correspond $J/M^2$: $0.9, 0.5$, and $0$
  respectively.  The solid line (outermost) is $J=0$.  When $J$ is
  increased, the horizon shrinks.}
\label{fig:rdbh_axi_ah}
\end{figure}
Fig.~\ref{fig:rdbh_axi_ah} shows the coordinate location of the
apparent horizon for the axisymmetric rotating cases with parameters
$(Q_0,J)=(1,J)$.  The solid (outermost) line corresponds to
$J=0$. When $J$ is increased, the AH tends to shrink.  For high spin
cases, the horizon will shrink.  This behavior is similar to the
location of the AH of Kerr.  Thus, the AH's should be on the throat
for the above parameter sets.

\begin{figure}[t]
  \epsfxsize=10cm
  \epsfysize=10cm
  \hspace{3cm}
  \epsfbox{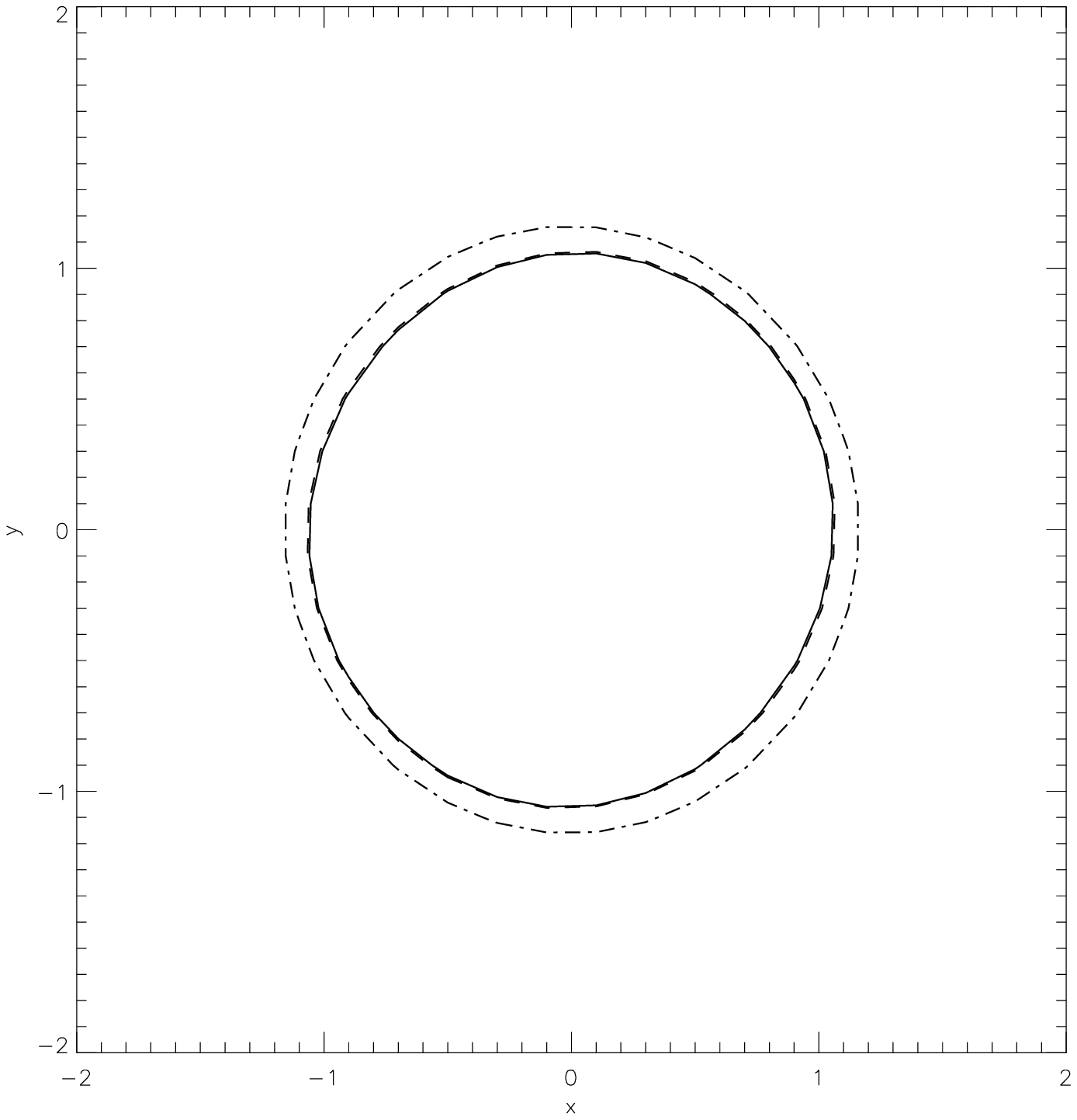}
  \caption{The coordinate location of the apparent horizon of
    non-axisymmetric cases for various total momentum parameter $J$:
    $75$, $10$, and $0$ which correspond $J/M^2$: $0.9, 0.5$, and $0$
    respectively.  Contradictory to axisymmetry cases, the outermost
    one is high spin:$J=75$.}
  \label{fig:rdbh_nonaxi_ah}
\end{figure}
Fig.~\ref{fig:rdbh_nonaxi_ah} shows the coordinate location of
the apparent horizon for non-axisymmetric rotating cases with parameters
the same as in the axisymmetric case: $(Q_0,J)=(1,J)$.  The
dot and dashed line (outermost) corresponds to $J=75\, (J/M^2=0.9)$.
For the rapidly spinning cases, horizons are expected to expand.  This
means that compared with axisymmetric cases, when we increase the
Brill wave amplitude, the AH horizons will leave the throat easily.

When the apparent horizon is on the throat, it is a coordinate
sphere.  It is interesting, however, to study how distorted it is in
the physical space.  One measure of this distortion is
the ratio of the polar to equator circumference:
$C_r=C_P/C_E$.  The polar circumference, $C_P$, is the proper length
of the circumference which goes through the poles along a line of
constant $\phi$, and the equatorial circumference, $C_E$, is the
proper length of equator, defined as $\theta=\pi/2$.  In axisymmetry,
this ratio is independent of $\phi$, and is a single number which
characterizes a data set once one has established the above
definitions. In this case, if $C_P/C_E=1$, the horizon is spherical,
if $C_P/C_E<1$ it is oblate, and if $C_P/C_E>1$ it is prolate.
In 3D, one can establish analogous definitions, although in this case
$C_P$, and therefore $C_r$, is a function of $\phi$.

\begin{figure}[t]
\epsfxsize=10cm
\epsfysize=10cm
\hspace{3cm}
   \epsfbox{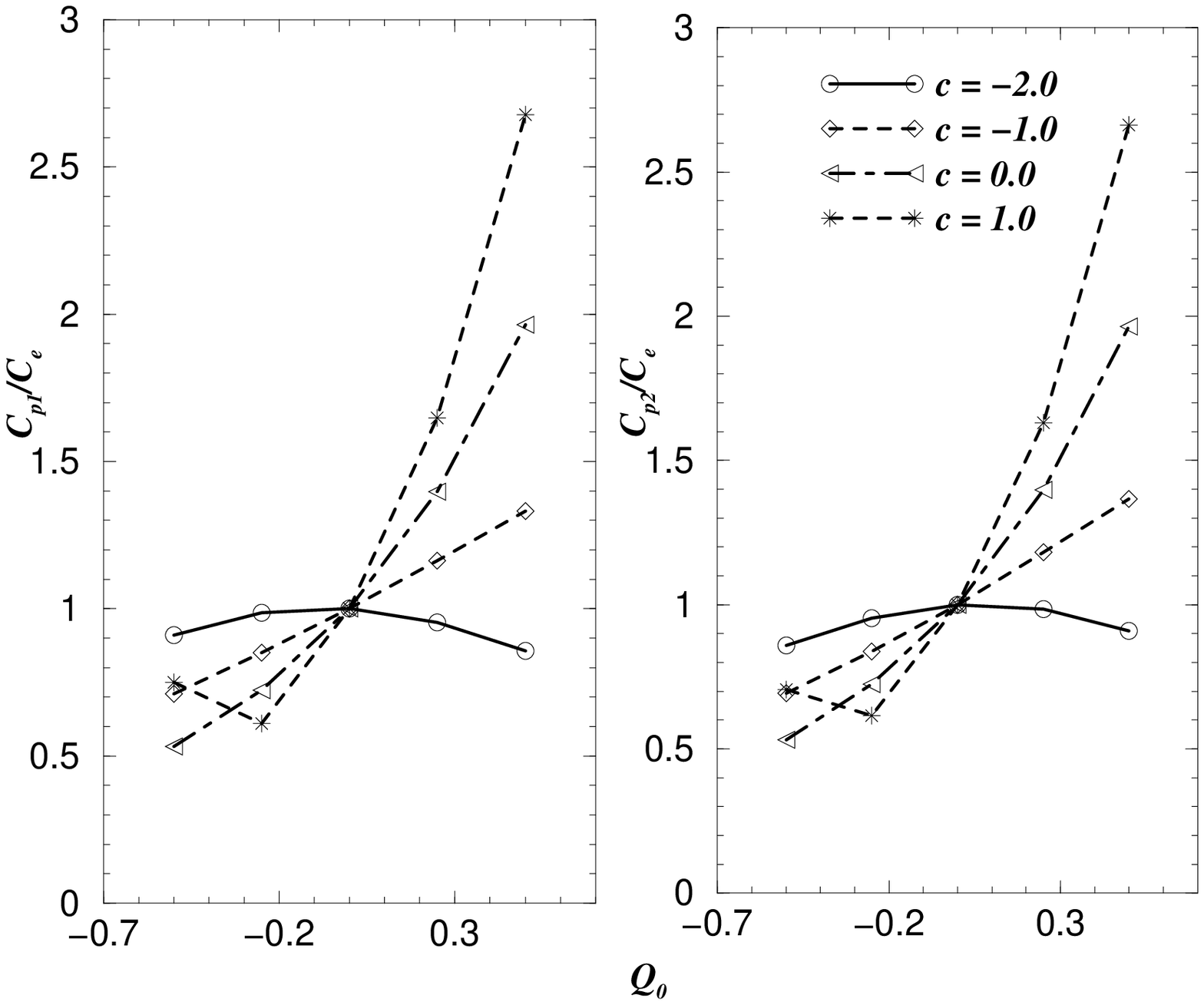} 
\caption{The ratio $C_P/C_E$ of non-rotating case for the parameter sets
    $(Q_0,\eta_0,\sigma,c) = (Q_0,0,1,c)$.  $C_{P1}$ is the value of
    $C_P$ at $\phi=0$, and $C_{P2}$ is the value of $C_P$ at
    $\phi=\pi/2$.}
  \label{fig:dbh_cp-ce}
\end{figure}
Fig.~\ref{fig:dbh_cp-ce} shows the ratio $C_P/C_E$ for two values of $\phi$
for non-rotating
cases with parameters $(Q_0,\eta_0,\sigma,c)=(Q_0,0,1,c)$.  $C_{P1}$
is the value of $C_P$ at $\phi=0$, and $C_{P2}$ is the value of $C_P$
at $\phi=\pi/2$.  We observe that highly distorted features appear when
$c>0$ and $Q_0>0$.

For distorted rotating black holes, we study the same parameter sets
as above: $(Q_0,\eta_0,\sigma,J,m) = (1,0,1,J,m)$.
\begin{figure}[t]
  \epsfxsize=10cm
  \epsfysize=10cm
  \hspace{3cm}
  \epsfbox{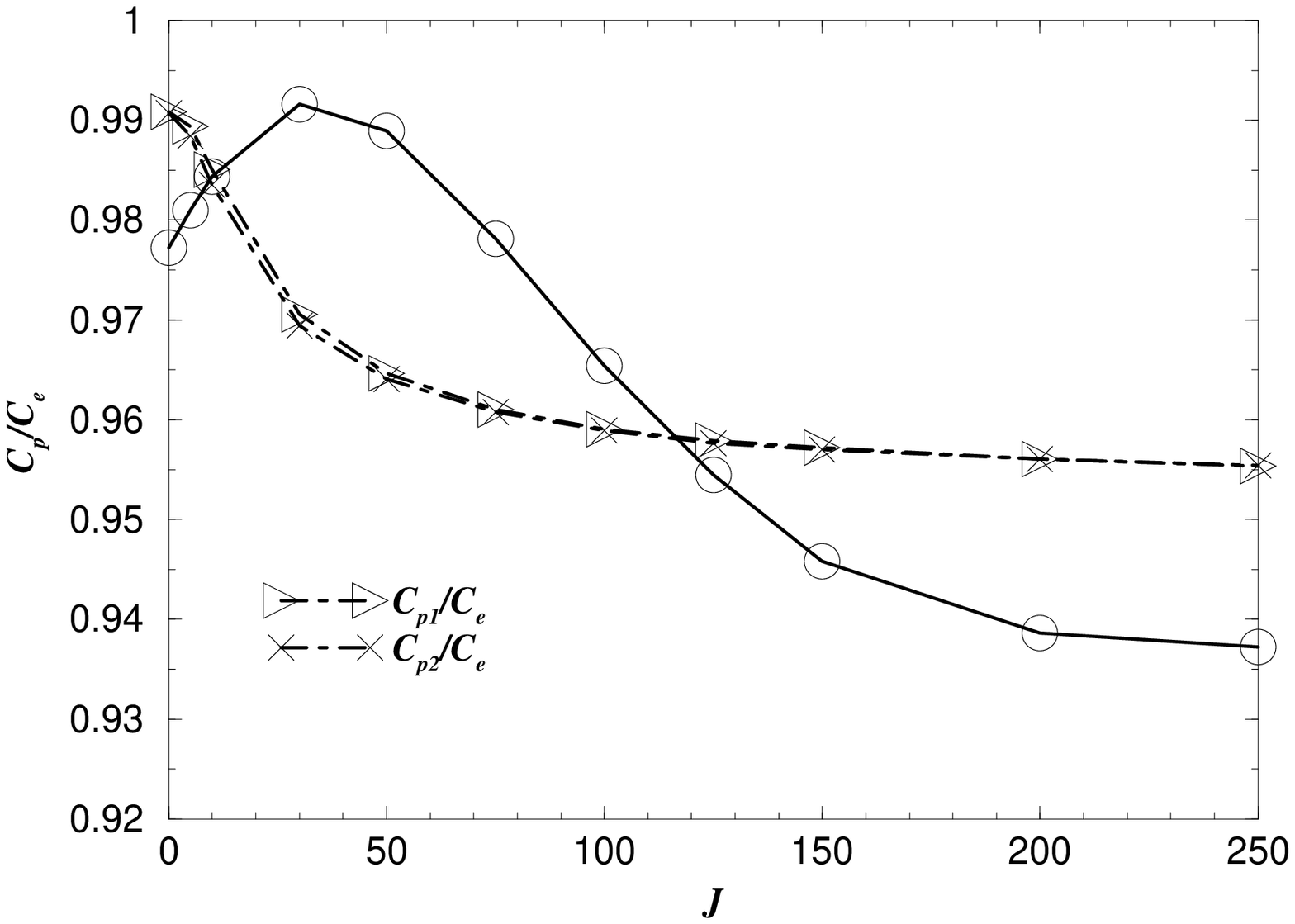}
  \caption{The ratio $C_P/C_E$ for axi- and non-axisymmetric rotating cases.
    The parameter sets are $(Q_0,\eta_0,\sigma,J)=(1,0,1,J)$.  For
    non-axisymmetric cases, we have $C_{P1}$ which is the value of
    $C_P$ at $\phi=0$, and $C_{P2}$ which is the value of $C_P$ at
    $\phi=\pi/2$.  The solid line shows the axisymmetric case,
    and dashed lines show non-axisymmetric cases.  High spin
    axisymmetric cases are slightly more oblate than the other cases.}
  \label{fig:rdbh_cp-ce}
\end{figure}
Fig.~\ref{fig:rdbh_cp-ce} shows $C_r$ as a function of $J$ for
$m=0$ and $m=2$.
Although the distortion patterns are different for
axisymmetric and non-axisymmetric cases, both cases are nearly
spherical for the range of parameters studied. High spin axisymmetric
cases are slightly more oblate than the other cases.  Also,
for the non-axisymmetric cases, $C_{P1}$ and $C_{P2}$ are indistinguishable.

\subsection{Maximum Radiation Loss ($MRL$)}
\label{subsec:mrl}

Another quantity of interest which is related to the apparent horizon
is the maximum radiation loss ($MRL$).  The
second law of black hole thermodynamics states that the area of the event
horizon of a black hole, $A_{EH}$, cannot decrease in
time.  One can define a quantity associated with the area of black
hole known as the irreducible mass,
$M_{irr}$~\cite{Christodoulou70,Cook90},
\begin{equation}
  \label{eqn:irr_mass}
  M_{irr} = \sqrt{\frac{A_{EH}}{16 \pi}}.
\end{equation}
By analogy to this quantity, one can define the mass of the apparent
horizon given its area, $A_{AH}$,
\begin{equation}
  M_{AH} = \sqrt{\frac{A_{AH}}{16 \pi}}.
\end{equation}
More generally, for a rotating black hole, one can define the apparent
horizon mass as
\begin{equation}
  \label{eqn:ah_mass}
  M^2_{AH} = \frac{A_{AH}}{16\pi} + \frac{4 \pi J^2}{A_{AH}},
\end{equation}
where $J$ is the total angular momentum.  

One defines the radiation efficiency $RE$ as the difference between
the ADM mass of a spacetime and its irreducible mass in its final
state:
\begin{equation}
  RE = \frac{M - M_{irr,f}}{M}.
\end{equation}
However, by second law of black hole thermodynamics, the final
irreducible mass must be greater than or equal to the initial
irreducible mass.  Thus,
\begin{equation}
RE \le \frac{M-M_{irr,i}}{M}.
\end{equation}
Because one needs to perform an evolution to find the location of the
event horizon on any slice, we can not compute the above fraction,
even though it is for quantities defined on our data set.  However, for 
our initial data sets, we know that the apparent horizon is the
outer-most minimal area surface.  It can be shown that if an apparent
horizon exists, it must lie inside or coincident with an event
horizon~\cite{Hawking73}.  Thus the area of the event horizon on our
initial slice must be greater than or equal to the area of the
apparent horizon.  Then we can define a quantity we call the maximum
radiation loss ($MRL$) by using the ADM mass $M$:
\begin{equation}
MRL = \frac{M-M_{AH,i}}{M} \ge RE.
\label{eqn:mrl}
\end{equation}
The fraction above is an upper bound on the radiation efficiency.
Therefore, we know that during an actual evolution, the total energy
radiated has to be less than this amount.  This will be a guide to
choosing interesting initial data sets for evolution in the future.

\begin{figure}[t]
\epsfxsize=10cm
\epsfysize=10cm
\hspace{3cm}
 \epsfbox{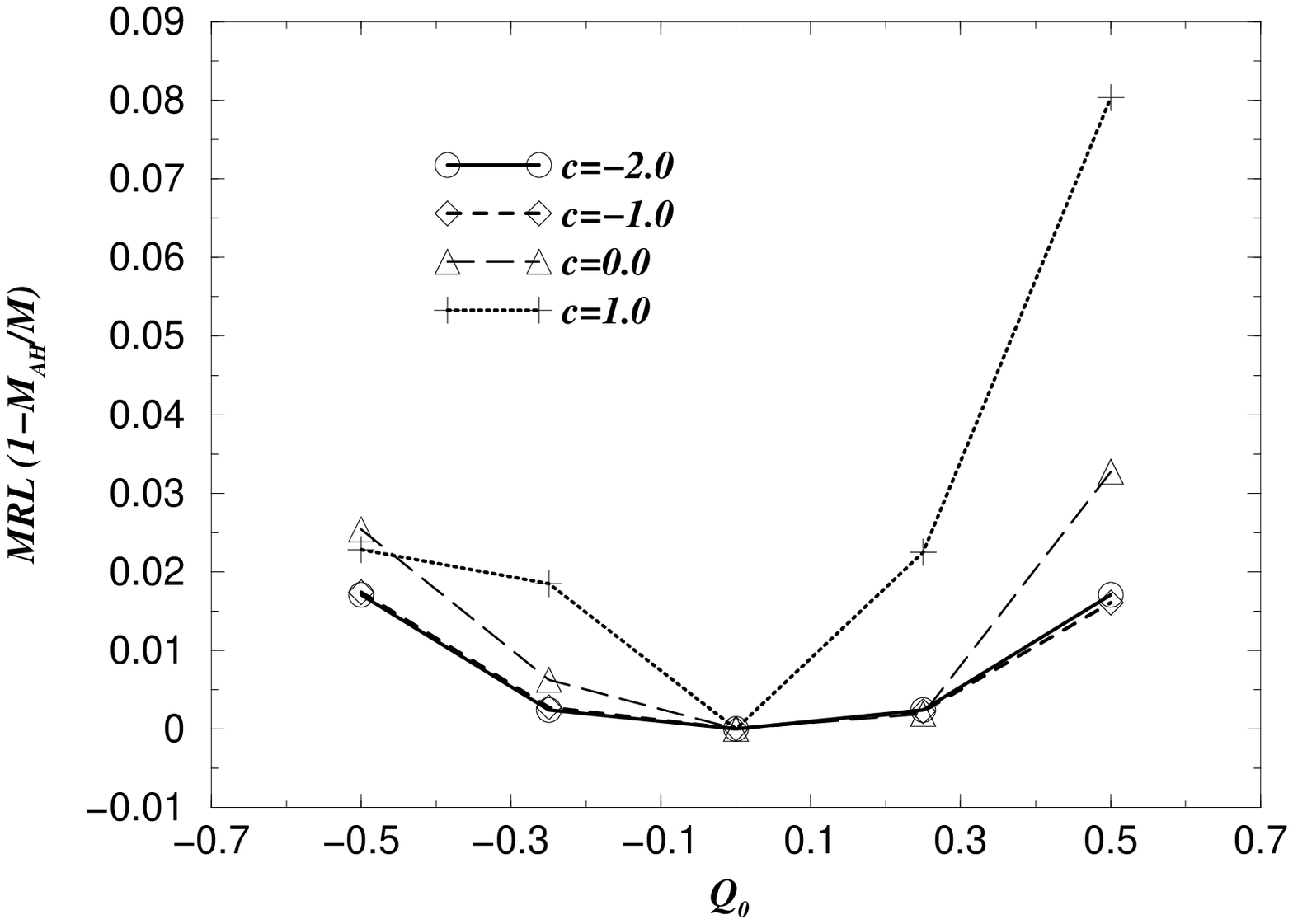}
  \caption{ We show the $MRL$ for the initial data set:
    $(Q_0,\eta_0,\sigma,c) = (Q_0,0,1,c)$.  The efficient $MRL$s are for 
    the positive amplitude and azimuthal factor.}
  \label{fig:dbh_mrl}
\end{figure}
Fig.~\ref{fig:dbh_mrl} shows the $MRL$ for non-rotating cases as a
function of $Q_0$ and $c$ for the family of initial data sets
$(Q_0,\eta_0,\sigma,c) = (Q_0,0,1,c)$.  The largest $MRL$ is about
$10$\% of the total mass.  The sets within this family
with the largest $MRL$s are those with both positive $Q_0$ and $c$,
indicating that these might be interesting data sets to evolve.

\begin{figure}[t]
\epsfxsize=10cm
\epsfysize=10cm
\hspace{3cm}
    \epsfbox{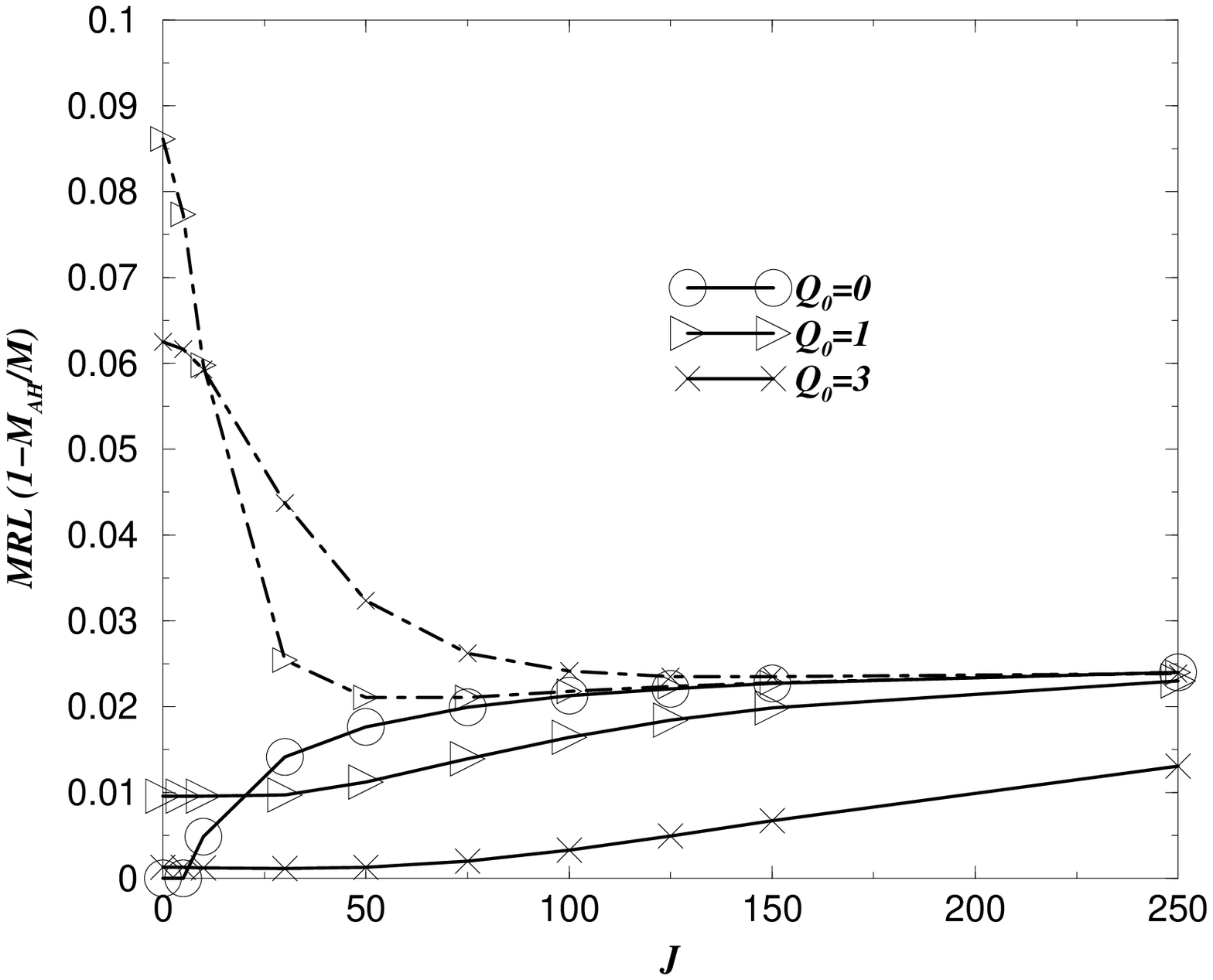} 
    \caption{We show the $MRL$ for rotating cases.  The initial data sets
      are $(Q_0,\eta_0,\sigma,J) = (Q_0,0,1,J)$.  The solid
      lines show the axisymmetry cases, and the dashed line shows the
      non-axisymmetry cases.}
    \label{fig:rdbh_mrl}
\end{figure}
Fig.~\ref{fig:rdbh_mrl} shows the $MRL$ for rotating cases as a function
of $J$ and $Q_0$ for the family of initial data sets
$(Q_0,\eta_0,\sigma,J) = (Q_0,0,1,J)$. For the axisymmetric cases, when
the initial amplitude is larger the $MRL$ tends to decrease, while
for non-axisymmetric cases the trend is reversed for most values of $J$.  Also, 
for the axisymmetric cases, the $MRL$ increases as $J$ becomes larger, but for
non-axisymmetric cases the $MRL$ decreases.  As we saw for the ADM masses,
when $J$ is increased,
the $MRL$ for both classes of initial data approaches the pure Bowen and York
values.

Furthermore, we can address an important question.  Penrose proposed
a criterion for a spacetime which, if violated, would indicate that Cosmic
Censorship would be violated~\cite{Penrose73}.  The criterion, the
so-called Penrose inequality, is simply that the irreducible apparent
horizon mass must be less than or equal to the ADM mass.  This would
result in a negative value of the $MRL$.  Using the Brill wave plus
black hole initial data, violations of the Penrose inequality were looked for
in Ref.~\cite{Bernstein93a,Brandt94a,Bernstein94a}, but none were found.
No violations were observed in either our non-rotating or rotating initial
data sets.

%
%

\section{CONCLUSIONS}
\label{sec:conc}

We have extended Bernsteins's axisymmetric, distorted black hole data
sets to full 3D by giving them a particular azimuthal dependence in the
Brill wave function $q$.  For rotating initial data, we have
generalized axisymmetric distorted rotating black hole by allowing
the conformal extrinsic curvature to have an azimuthal dependence.

From their physical properties, such as their ADM mass distortion of
horizons and $MRL$, non-rotating axisymmetric cases are
consistent with what has been observed by Bernstein's.  In particular,
Bernstein's results are recovered when the azimuthal parameter $c$ is set
to zero. However, rotating data
sets are, by construction, different from those of Brandt and Seidel.
For those initial data sets, we have seen that one of
distortion parameters, $Q_0$, strongly effects the physical
properties when $J$ is low, as we showed in Section~\ref{sec:survey}.
However, as $J$ is increased, physical properties are approaching the
pure Bowen-York data.  For further investigation of higher
amplitude and spin cases for full 3D initial data as in
Ref.~\cite{Dain:2002ee}, since we are using uni-grid Cartesian coordinate
for all physical quantities except for the ADM mass, we need larger
grid size and higher resolutions by using such as adaptive mesh
refinement.

In future papers we would like to address the radiation of angular
momentum by evolving non-axisymmetric, distorted, rapidly-rotating black
holes.  Direct comparisons of radiative energy were done for axisymmtric
and non-rotating initial
data~\cite{Allen97a,Allen98a,Baker99a,Baker96a}.  The close limit
approximation studies are addressing radiated angular
momentum~\cite{Khanna99a,Khanna:2000dg}.  However, as mentioned in
Ref.~\cite{Alcubierre00b}, a direct comparison with full numerical
simulations is under investigation.  By using the Lazarus
method~\cite{Baker00b,Baker:2001sf}, Ref.~\cite{Alcubierre00b}, one shows that
initial data for grazing collisions cannot be mapped into the
perturbative method.  Although our non-axisymmetric, distorted rapidly
rotating black holes are not necessarily astrophysically relevant,
such analysis will provide an example of the usefulness of
perturbation theory as an interpretive tool for understanding the
dynamics produced in full nonlinear evolutions.  

\acknowledgements 
We would like to thank our colleagues at AEI, Penn State, and TAC
, Miguel Alcubierre, John Baker, Bernd Br\"ugmann, Manuela
Campanelli, Carlos O. Lusto, and Igor D Novikov. Special thanks to Sergio 
Dain for various discussion and comments.  K.~C. acknowledges support from
Microsoft. Calculations were performed at the AEI on
an SGI Origin 2000 and the Leibniz-Rechenzentrum on SR8000-F 
supercomputers.

%
%


\begin{thebibliography}{10}


\bibitem{Bernstein93a}
D. Bernstein, Ph.D. thesis, University of Illinois Urbana-Champaign, 1993.

\bibitem{Brandt94a}
S. Brandt and E. Seidel, Phys. Rev. D {\bf 54},  1403  (1996).

\bibitem{Abramovici92}
A.~A. Abramovici {\it et~al.}, Science {\bf 256},  325  (1992).

\bibitem{Hough94b}
J. Hough,   (1994), prepared for Edoardo Amaldi Meeting on Gravitational Wave
  Experiments, Rome, Italy, 14-17 Jun 1994.

\bibitem{Flanagan97a}
\'{E}anna \'{E}.~Flanagan and S.~A. Hughes, Phys. Rev. D {\bf 57},  4535
  (1998).

\bibitem{Damour:2001bu}
T. Damour, P. Jaranowski, and G. Sch{\"a}fer, Phys. Lett. {\bf B513},  147
  (2001).

\bibitem{Damour:2000ni}
T. Damour, P. Jaranowski, and G. Sch{\"a}fer, Phys. Rev. {\bf D63},  044021
  (2001).

\bibitem{Damour:2000we}
T. Damour, P. Jaranowski, and G. Sch{\"a}fer, Phys. Rev. {\bf D62},  084011
  (2000).

\bibitem{Jaranowski98a}
P. Jaranowski and G. Sch\"afer, Phys. Rev. D {\bf 57},  7274  (1998).

\bibitem{Blanchet95}
L. Blanchet {\it et~al.}, Phys. Rev. Lett. {\bf 74},  3515  (1995).

\bibitem{Price94a}
R.~H. Price and J. Pullin, Phys. Rev. Lett. {\bf 72},  3297  (1994).

\bibitem{Tichy02}
{W. Tichy , B. Br\"ugmann, M. Campanelli, and P. Diener},
  (2002), gr-qc/0207011.

\bibitem{Anninos93b}
P. Anninos {\it et~al.}, Phys. Rev. Lett. {\bf 71},  2851  (1993).

\bibitem{Brandt00}
S. Brandt {\it et~al.}, Phys. Rev. Lett. {\bf 85},  5496  (2000).

\bibitem{Alcubierre00b}
M. Alcubierre {\it et~al.}, Phys. Rev. Lett. {\bf 87},  271103  (2001),
  gr-qc/0012079.

\bibitem{Abrahams92a}
A. Abrahams {\it et~al.}, Phys. Rev. D {\bf 45},  3544  (1992).

\bibitem{Brandt94c}
S. Brandt and E. Seidel, Phys. Rev. D {\bf 52},  870  (1995).

\bibitem{Alcubierre01a}
M. Alcubierre {\it et~al.}, Phys. Rev. D {\bf 64},  R61501  (2001).


\bibitem{Brandt97c}
S. Brandt, K. Camarda, and E. Seidel,  in {\em Proceedings of the 8th Marcel
  Grossmann Meeting on General Relativity}, edited by T. Piran (World
  Scientific, Singapore, 1999), pp.\ 741--743.

\bibitem{Allen97a}
G. Allen, K. Camarda, and E. Seidel,   (1998), gr-qc/9806014, submitted to
  Phys.\ Rev.\ D.

\bibitem{Allen98a}
G. Allen, K. Camarda, and E. Seidel,   (1998), gr-qc/9806036, submitted to
  Phys.\ Rev.\ D.

\bibitem{Baker99a}
J. Baker {\it et~al.}, Phys. Rev. D {\bf 62},  127701  (2000), gr-qc/9911017.

\bibitem{Alcubierre01b}
M. Alcubierre {\it et~al.},   (2001), in preparation.

\bibitem{Camarda97a}
K. Camarda, Ph.D. thesis, University of Illinois at Urbana-Champaign, Urbana,
  Illinois, 1998.

\bibitem{York79}
J. York,  in {\em Sources of Gravitational Radiation}, edited by L. Smarr
  (Cambridge University Press, Cambridge, England, 1979).

\bibitem{Brill59}
D.~S. Brill, Ann. Phys. {\bf 7},  466  (1959).

\bibitem{Bondarescu02}
M. Bondarescu, M. Alcubierre, and E. Seidel, Class.Quant.Grav. {\bf 19},  375
  (2002).

\bibitem{Brandt94b}
S. Brandt and E. Seidel, Phys. Rev. D {\bf 52},  856  (1995).

\bibitem{Bowen80}
J. Bowen and J.~W. York, Phys. Rev. D {\bf 21},  2047  (1980).

\bibitem{Dain02a}
S. Dain,   (2002), gr-qc/0207090.

\bibitem{Dain01a}
S. Dain and H. Friedrich, Comm. Math. Phys. {\bf 222},  569  (2001).

\bibitem{Arnowitt62}
R. Arnowitt, S. Deser, and C.~W. Misner,  in {\em Gravitation: An Introduction
  to Current Research}, edited by L. Witten (John Wiley, New York, 1962), pp.\
  227--265.

\bibitem{Omurchadha74}
N. O'Murchadha and J. York, Phys. Rev. D {\bf 10},  2345  (1974).

\bibitem{Dain:2002ee}
S. Dain, C.~O. Lousto, and R. Takahashi, Phys. Rev. D {\bf 65},  104038
  (2002).

\bibitem{York89}
J. York,  in {\em Frontiers in Numerical Relativity}, edited by C. Evans, L.
  Finn, and D. Hobill (Cambridge University Press, Cambridge, England, 1989),
  pp.\ 89--109.

\bibitem{Gibbons72}
G. Gibbons, Commun. Math. Phys. {\bf 27},  87  (1972).

\bibitem{Alcubierre98b}
M. Alcubierre {\it et~al.}, Class. Quantum Grav. {\bf 17},  2159  (2000).

\bibitem{Christodoulou70}
D. Christodoulou, Phys. Rev. Lett. {\bf 25},  1596  (1970).

\bibitem{Cook90}
G. Cook, Ph.D. thesis, University of North Carolina at Chapel Hill, Chapel
  Hill, North Carolina, 1990.

\bibitem{Hawking73}
S.~W. Hawking,  in {\em Black Holes}, edited by C. DeWitt and B.~S. DeWitt
  (Gordon and Breach, New York, 1973), pp.\ 1--55.

\bibitem{Penrose73}
R. Penrose, Ann. N.Y. Acad. Sci. {\bf 224},  125  (1973).

\bibitem{Bernstein94a}
D. Bernstein, D. Hobill, E. Seidel, and L. Smarr, Phys. Rev. D {\bf 50},  3760
  (1994).

\bibitem{Baker96a}
J. Baker {\it et~al.}, Phys. Rev. D {\bf 55},  829  (1997).

\bibitem{Khanna99a}
G. Khanna {\it et~al.}, Phys. Rev. Lett. {\bf 83},  3581  (1999).

\bibitem{Khanna:2000dg}
G. Khanna, R. Gleiser, R. Price, and J. Pullin, New Jour. Phys. {\bf 2},  3
  (2000).

\bibitem{Baker00b}
J. Baker, B. Br\"ugmann, M. Campanelli, and C.~O. Lousto, Class. Quantum Grav.
  {\bf 17},  L149  (2000).

\bibitem{Baker:2001sf}
J. Baker, M. Campanelli, and C.~O. Lousto, Phys. Rev. {\bf D65},  044001
  (2002).

\end{thebibliography}
\end{document}